\documentclass[conference]{IEEEtran}
\IEEEoverridecommandlockouts
\usepackage{cite}

\usepackage{amsmath,amssymb,amsfonts,mathtools,bm}
\usepackage{amsthm}
\usepackage{algorithm}
\usepackage{algorithmic}
\usepackage{graphicx}
\usepackage{textcomp}
\usepackage{xcolor,float}
\usepackage{tabularx}
\usepackage{textcomp}
\usepackage{diagbox}
\usepackage{svg}
\usepackage{booktabs}
\usepackage[colorlinks,
            linkcolor=blue,
            anchorcolor=blue,
            citecolor=blue]{hyperref}

\allowdisplaybreaks
\renewcommand{\baselinestretch}{1}

\begin{document}
    
\title{{Interpretable and Secure Trajectory Optimization for UAV-Assisted Communication}}

\author{
\IEEEauthorblockN{
Yunhao Quan\IEEEauthorrefmark{1},
Nan Cheng\IEEEauthorrefmark{1},
Xiucheng Wang\IEEEauthorrefmark{1},
Jinglong Shen\IEEEauthorrefmark{1},
Longfei Ma\IEEEauthorrefmark{1},
and Zhisheng Yin\IEEEauthorrefmark{2}\\
}
\IEEEauthorblockA{
\IEEEauthorrefmark{1}School of Telecommunications Engineering,
Xidian University, Xi'an, China\\
\IEEEauthorrefmark{2}School of Cyber Engineering,
Xidian University, Xi'an, China\\
Email: \{qyh, xcwang\_1, jlshen, lfma\}@stu.xidian.edu.cn,, dr.nan.cheng@ieee.org, zsyin@xidian.edu.cn}}
    
\maketitle

\IEEEdisplaynontitleabstractindextext

\IEEEpeerreviewmaketitle

\begin{abstract}
Unmanned aerial vehicles (UAVs) have gained popularity due to their flexible mobility, on-demand deployment, and the ability to establish high probability line-of-sight wireless communication. As a result, UAVs have been extensively used as aerial base stations (ABSs) to supplement ground-based cellular networks for various applications. However, existing UAV-assisted communication schemes mainly focus on trajectory optimization and power allocation, while ignoring the issue of collision avoidance during UAV flight. To address this issue, this paper proposes an interpretable UAV-assisted communication scheme that decomposes reliable UAV services into two sub-problems. The first is the constrained UAV coordinates and power allocation problem, which is solved using the Dueling Double DQN (D3QN) method. The second is the constrained UAV collision avoidance and trajectory optimization problem, which is addressed through the Monte Carlo tree search (MCTS) method. This approach ensures both reliable and efficient operation of UAVs. Moreover, we propose a scalable interpretable artificial intelligence (XAI) framework that enables more transparent and reliable system decisions. The proposed scheme's interpretability generates explainable and trustworthy results, making it easier to comprehend, validate, and control UAV-assisted communication solutions. Through extensive experiments, we demonstrate that our proposed algorithm outperforms existing techniques in terms of performance and generalization. The proposed model improves the reliability, efficiency, and safety of UAV-assisted communication systems, making it a promising solution for future UAV-assisted communication applications.
\end{abstract}


\section{Introduction}
In recent years, unmanned aerial vehicles (UAVs) have become increasingly popular due to their maneuverability, on-demand deployment capabilities, and ability to establish line-of-sight (LOS) wireless communication links with high probability\cite{zhong2022}. Despite their advantages, the reliability and safety of UAV-assisted networks are largely dependent on the design of flying trajectories, power allocation, and guaranteed collision avoidance\cite{JSAC2018}. However, dynamic environments and unreliable wireless channels present significant challenges to the reliability of UAV services, making the control of UAV motion a critical issue for UAV-assisted communication systems.

Therefore, a considerable amount of research has been devoted to optimizing UAV trajectory. For instance, Zhan et al. \cite{ZHAN} have proposed an optimization method that maximizes the energy efficiency of sensor networks by optimizing the UAV trajectory. Similarly, in \cite{IOTJ}, the authors have optimized the trajectory, transmission power, and connection between UAVs and nodes to minimize the total transmission power in the system. Baek et al. \cite{TVT} have also investigated UAV trajectory and route design, modeling the UAV using a hovering flight model. Furthermore, researchers in \cite{TWC2019}, \cite{TWC2017}, and \cite{wang2019} have proposed UAV trajectory optimization schemes that aim to minimize energy consumption or maximize flying time. These works primarily focus on optimizing UAV trajectory and power allocation to reduce energy consumption or extend UAV flying time.

While optimizing UAV trajectory and power allocation is important, collision avoidance is equally crucial to ensure the safety and reliability of UAV networks. Jointly optimizing UAV trajectory, resource allocation strategy, and collision avoidance strategy is considered a potential solution. To this end, Yang et al. \cite{Yang} have proposed a deep deterministic policy gradient (DDPG) algorithm to jointly optimize UAV trajectory, resource allocation strategy, and interference strategy to maximize energy efficiency. Zhang et al. \cite{Zhang} have employed the deep Q-network (DQN) method to jointly design UAV transmission scheduling, power allocation, and trajectory optimization to maximize the system transmission rate. Liu et al. \cite{Liu} have utilized a multi-agent deep deterministic policy gradient (MADDPG) method to extract features through convolutional neural network (CNN) and jointly optimize UAV operational trajectory and collision avoidance. These works employ deep reinforcement learning methods to effectively address the issue of collision avoidance during UAV service operations. However, these methods lack interpretability, which can raise concerns about UAV safety and result in unnecessary legal disputes\cite{Guo2020}\cite{XAI1}. Therefore, ensuring the interpretability and reliability of decisions is essential in designing algorithms for UAV operations.

To enhance the performance of UAV-assisted communication networks, this paper proposes a joint optimization approach for trajectory and power allocation under collision avoidance conditions. Furthermore, an architecture based on explainable artificial intelligence (XAI) is presented to efficiently address this problem. The main contributions of this paper as follows.
\begin{enumerate}
    \item A scalable XAI framework is proposed to aid in UAV collision avoidance and trajectory optimization problems, which improves the credibility of decision-making processes.
    \item  To address the joint optimization problem of trajectory and power allocation under collision avoidance constraints, this paper decomposes the problem into two mutually constrained sub-problems. Firstly, a Double Dueling DQN (D3QN)-based method is used to solve the power allocation and service coordinate problem under collision avoidance constraints. Secondly, a Monte Carlo tree search (MCTS)-based method is utilized to solve the trajectory optimization problem under collision avoidance constraints.
    \item We conduct extensive experiments to evaluate the proposed joint optimization algorithm. The experimental results indicate that our algorithm outperforms existing algorithms in terms of both performance and generalization. In addition, the tree search method provides decision paths during the search process, which enhances the interpretability of our algorithm.
\end{enumerate}

The following paragraphs are organized as follows. In Section \ref{section:System Model and Problem Formulation}, we provide an overview of the system model, including its composition and function, as well as detailed instructions on how to use each module.  In 
Section \ref{section:UAV-Assisted Communication Methods}, 
we introduce the methods for solving various sub problems.
we present a large number of experiments and discussions in Section \ref{section:Simulation result}. Finally, we summarize the article in Section \ref{section:Conclusion}.

\section{System Model and Problem Formulation}
\label{section:System Model and Problem Formulation}

\begin{figure*}[h]
  \centering
  \includegraphics[width=1.0\linewidth]{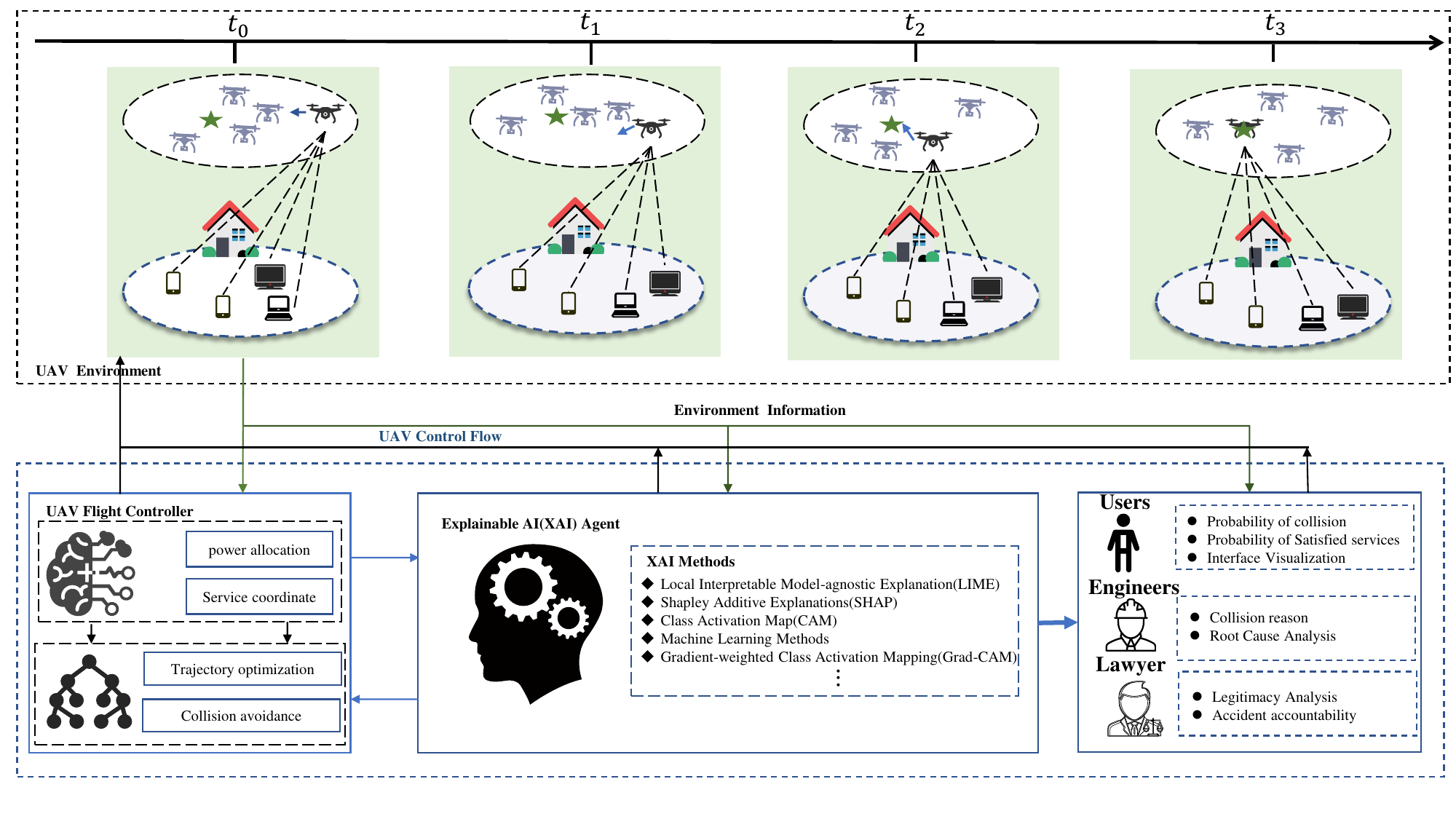}
   \vspace{-23pt}
  \caption {Scalable and interpretable artificial intelligence framework based on UAV-assisted communication.}
   \vspace{-9pt}
   \label{System Model}
\end{figure*}

As shown in Fig.~\ref{System Model}, we consider a UAV-assisted communication network consisting of multiple users and UAVs. The set of users served by the UAV is denoted as $k \in \mathbb{K}=\{1,2,3 \ldots K\}$, $h(t)$ represents the altitude of the UAV during flight. The distance between the UAV and the $k$-th user at time t is denoted as $d_{k}(t)$ that \begin{equation}
d_k\left(t\right)=\sqrt{h_u^2\left(t\right)+\left[x_u\left(t\right)-x_k\left(t\right)\right]^2+\left[y_u\left(t\right)-y_k\left(t\right)\right]^2}.\end{equation}
The average path loss between the UAV and the $k$-th user can be expressed as \begin{equation}
L_k(t)=P_{\mathrm{LoS}} \cdot L_{\mathrm{LoS}}+P_{\mathrm{NLoS}} \cdot L_{\mathrm{NLoS}.}
\end{equation}
Taking into account small-scale fading, the channel gain between the UAV and user $k$ at time $t$ can be calculated as\begin{equation}
g_{k}(t)=H_{k}(t) \cdot 10^{-L_{k}(t) / 10}.\end{equation}
$H_{k}(t)$ represents the channel fading coefficient between the UAV and user $k$ at time $t$, $v_{k}(t)$ serves as a performance metric, where $v_{k}(t)=1$ indicates that the UAV is serving $k$-th user, and $v_{k}(t)=0$ otherwise. $p_{k}(t)$ represents the power allocated to user $k$, and the data rate between user $k$ and the UAV can be represented as:\begin{equation}R_{(k)}(t)=B \log 2\left(1+\gamma_{(k)}(t)\right).\tag{4}\end{equation}\par
\begin{equation}\gamma_{k}(t)=\frac{v_{k}(t) g_{k}(t) P_{k}(t)}{\sum_{i=1, i \neq k}^{K} v_{k}(t) g_{k}(t) P_{k}(t)+\sigma_{k}(t)^{2}}.\tag{5}\end{equation}\par
$\sigma_{k}(t)$ represents additive Gaussian white noise, $\gamma_{k}(t)$ is the signal-to-noise ratio (SNR) of the channel between the $k$-th user and the UAV, B is the communication bandwidth of the UAV, Therefore, the overall rate of the system is given by 
\begin{equation}
{R}(t)= \sum_{k=1}^{K} \mathcal{R}_{(k)}(t).
\tag{6}
\end{equation}
The throughput of the system within time $T$ can be given by
\begin{equation}
{R}=\sum_{t=0}^{T} \mathcal{R}(t).
\tag{7}
\end{equation}
\par 
$\mathbf{H}=\left\{h(t),x(t),y(t),0 \leq t \leq T\right\}$ is the location coordinates of the UAV during service time, During the service time, The power allocated by UAV to each user is denoted by the variable $\mathbf{P}=\left\{p_{k}(t), 0 \leq t \leq T, k \in \mathbb{K}\right\}$. The variable $\mathbf{V}=\left\{v_{k}(t), 0 \leq t \leq T \right\}$ can be used to quantify the connectivity between users and the UAV.  $S_{t}$ expresses the number of steps the UAV has taken at a specific time $t$,  $S_{max}$ is the maximum number of steps that the UAV can fly. $\mathbf{U}=\left\{u_{k}(t), 0 \leq t \leq T \right\}$ represents the specific control action executed by the UAV during flight, such as its movement trajectory or any adjustments made to maintain a stable position in the air, ${\mathcal{C}}\left( H, U\right)$ is the collision statistics function. With the objective of maximizing system throughput and minimizing collision probability, subject to constraints on maximum power, spatial limitations, and Quality of Service (QoS) requirements, the problem of reliable service provision by UAVs can be formulated as follows.
\begin{align}
\max_{\mathbf{H}, \mathbf{V}, \mathbf{P}, \mathbf{U}} & \mathcal{G}=\sum_{t=0}^T (\mathcal{R}(t)-\mathcal{C}(\mathbf{H}, \mathbf{U})),\tag{8}
\label{equ8}\\
\text{s.t.~~~}
& h_{\min}\leq h(t)\leq h_{\max},\forall t\in[0, T],\tag{8a} \\
& x_{\min}\leq x(t)\leq x_{\max},\forall t\in[0, T],\tag{8b}\\
& y_{\min}\leq y(t)\leq y_{\max},\forall t\in[0, T],\tag{8c}\\
& \sum_{k \in \mathbb{K}} v_{k}(t) P_{k}\leq P_{\max},\forall t\in[0, T],\forall k\in\mathbb{K},\tag{8d}\\
& S_{t}\leq S_{\max},\forall t\in[0, T],\tag{8e}\\
& R_{k}(t)\geq R_{\mathrm{Qos}},\forall t\in[0, T],\forall k\in\mathbb{K}.\tag{8f}
\end{align}

It is worth noting that the optimization problem described above is a mixed exponential non-convex problem, which is known to be an NP hard problem. Moreover, in the scenario under consideration, both large-scale fading and small-scale fading are dependent on the instantaneous position of the UAV and users, making it difficult to solve the optimization problem using traditional optimization methods.  sub-problem decomposition and reinforcement learning have proven to be effective methods for dealing with complex control problems in high-dimensional continuous spaces. In the next section, we will adopt the idea of sub-problem decomposition to decompose the aforementioned problem and solve it using reinforcement learning and MCTS methods. Furthermore, as shown in Fig.~\ref{System Model}, a corresponding XAI framework will be designed.

\section{UAV-Assisted Communication Methods}
\label{section:UAV-Assisted Communication Methods}
Inspired by the idea of sub problem decomposition, the original problem was decomposed into two sub problems: power allocation and coordinate solving, as well as trajectory optimization and collision avoidance to reduce problem complexity.
\subsection{coordinate and power allocation}
UAV service coordinate solving and power allocation problems can be expressed as
\begin{align}
\max _{\mathbf{H}, \mathbf{V}, \mathbf{P}} & \mathcal{R}=\sum_{t=0}^{T} \mathcal{R}(t),\tag{9}
\label{equ8}\\
\text {s.t.~~}
  & h_{\min } \leq h(t) \leq h_{\max },\forall t \in [0, T],\tag{9a} \\
& x_{\min } \leq x(t) \leq x_{\max },  \forall t \in [0, T], \tag{9b}\\
& y_{\min } \leq y(t) \leq y_{\max },  \forall t \in [0, T], \tag{9c}\\
& \sum_{k \in \mathbb{K}} v_{k}(t) P_{k} \leq P_{max}, \forall t \in [0, T],  \forall k\in\mathbb{K}, \tag{9d}\\
& R_{k}(t) \geq R_{\mathrm{Qos}}, \forall t \in [0, T],  \forall k\in\mathbb{K}.\tag{9e}
\end{align}

The D3QN reinforcement learning algorithm is utilized to solve the problem, which adopts two neural networks to fit state and action values, and an extra layer to estimate the advantage values of each action. Specifically, the Q-value for each action at each time step can be calculated by leveraging this Q-value and the difference in the average advantage value of other actions. For a more detailed description of the algorithm flow, please refer to Algorithm \ref{Alg1}. 

$\bullet$ \textbf{Action Space:}
The action space comprises the UAV's moving direction and the power allocated to each user, expressed as a vector of size $K\times6$. UAV has seven available movement options: move left, move right, move forward, move backward, ascend, descend, or remain stationary. Simultaneously, the summation of all power allocation values must not exceed the power constraint.

$\bullet$ \textbf{State:}
The state space consists of the three-dimensional position of the UAV and the channel gain between the drone and the users. 

$\bullet$ \textbf{Reward:}
To maximize the overall throughput, we design the reward function as follows, $\lambda$ represents the penalty factor.
\begin{equation}
    R=\frac{\mathcal{R}(t)}{2^{\lambda}}.
    \label{equ14}
    \tag{10}
\end{equation}

In the D3QN model, the connected UAV and users first input abstract state information into the evaluation network to determine the optimal action. Next, the reward value is calculated, and the corresponding action is executed in the environment. Once a UAV-terminal user pair completes the service, we calculate the data rate for that time period. 
 \begin{algorithm}[!h]
	\caption{D3QN  algorithm for UAV service coordinates solution}
    \label{Alg1}
	\begin{algorithmic}[1]
	\FOR{each episode }
    \STATE { Initialize initial positions of UAV and users}
    \STATE {Initialize the network parameter $\theta$}
    \STATE {Update $\epsilon$ in action policy}
    \FOR{each step $t_{0} \leq t \leq t_{0}+T_{r}$ }
    \STATE{Calculate $g_{k}(t)$}
    \STATE{Generate state abstraction array $s$}
    \STATE{Choose A according to action policy and $Q(s, a, \theta)$}
    \STATE{Take action $a$,observe $r$ and $s^{'}$}
    \STATE{Store $D=(s,s^{'},r,a)$}
    \STATE{Sample random mini-batch of transitions $(s_j,a_j,r_j,s_{j+1})$ from $D$}
    \STATE {Set $y_j = r_j + \gamma \max_{a'} \hat{Q}(s_{j+1},a';\theta^{-})$}
    \STATE {Update the action-value function using gradient descent: $\Delta\theta = \alpha(y_j - Q(s_j,a_j;\theta))\nabla_\theta Q(s_j,a_j;\theta)$}
    \ENDFOR
    \ENDFOR
    
	\end{algorithmic}
\end{algorithm}
 \subsection{trajectory optimization and collision avoidance}
 The trajectory optimization and collision avoidance problems can be calculated as 
 \begin{align}
\min _{\mathbf{H}, \mathbf{U}} & \mathcal{C}=\sum_{t=0}^{T} \mathcal C{( H, U)},\tag{11}
\label{equ8}\\
\text {s.t.~}
  & h_{\min } \leq h(t) \leq h_{\max },\forall t \in [0, T],\tag{11a} \\
& x_{\min } \leq x(t) \leq x_{\max },  \forall t \in [0, T], \tag{11b}\\
& y_{\min } \leq y(t) \leq y_{\max },  \forall t \in [0, T], \tag{11c}\\
& S_{t} \leq S_{max}, \forall t \in [0, T], \tag{11d}
 \end{align}

To address this issue, we treat it as a Markov Decision Process (MDP) problem. ($p_{x}^{(k)}$,$p_{y}^{(k)}$) and ($v_{x}^{(k)}$,$v_{y}^{(k)}$) respectively represent the position coordinates and velocity of the $k$th intruder. ($o_{x},o_{y}$) and ($v_{x},v_{y}$) are the position coordinates and velocity of the ownership. $A_{\psi}$ and $A_{\phi}$ is the heading angle and tilt angle of the ownership. MCTS method is utilized to tackle the aforementioned problem.

\textbf{Action space:} 
At the outset of each time step, the target aircraft chooses to adjust its tilt angle and acceleration at a certain rate. $\mathcal{A}_{\phi}$ represents the directional action space,which consists of three actions: left turn, straight, and right turn. $\mathcal{A}_{a}$ represents the acceleration action space,which consists of three actions: Speed up, slow down, and constant speed.

\textbf{Termination state:} For safety reasons, we define the minimum collision distance between two UAVs as $d_{min}$. If the distance between two UAVs is less than $d_{min}$, it can be considered a collision. the termination state of the entire process can be divided into three types:
\begin{enumerate}
    \item The distance between the intruder and the ownership is less than $d_{min}$ (can be considered as a collision).
    \item The ownership out of the map we defined or cannot reach the goal within the specified steps (time out).
    \item The ownership successfully reaches the goal (goal).
\end{enumerate}\par
In MCTS, the nodes of the search tree correspond to states in the state space. Meanwhile, the leaf nodes encompass all possible next nodes (states) that can result from different actions performed by the current node (state). Given that each time step involves 9 action spaces, each node may have up to 9 leaf nodes.\par
The MCTS algorithm selects actions by forward searching the search tree. Each edge $(s,a)$ in the search tree stores an action value $Q(s,a)$ and its number of visits $N(s,a)$. The tree is traversed through simulation starting from the root node (initial state). The MCTS algorithm can be divided into four steps:
\begin{enumerate}
    \item The ownership will select the leaf node with the highest value according to Equation (\ref{equ15}), which maximizes the sum of the average action value and the uncertainty reward.
   \begin{equation}
        U C T=\bar{X}_{j}+2 C \sqrt{\frac{2 \ln n}{n_{j}}}.
        \label{equ15}
        \tag{12}
    \end{equation}
    The variable $\bar{X}_{j}$ approximately represents the state-action value of the child node, $U C T=2 C \sqrt{\frac{2 \ln n}{n_{j}}}$ known as the exploration term,  $n_{j}$ represents the number of times child node $j$ has been visited, and $n$ represents the number of times the parent node has been visited. $C$ is a constant that balances exploration and exploitation. If multiple child nodes have the same maximum value, the leaf node will be randomly selected. If a child node has never been visited, it will be prioritized, ensuring that each leaf node is visited at least once.
    \item When the ownship enters a node (state) that it has not yet visited, a new node is created as a child node of the parent node (i.e., the previous state) in the search tree. The visit count of this new node is set to 1, and the cumulative reward value is initialized to 0.
    \item Ordinarily, a large number of iterations are required by the conventional approach to reach a termination state by following a random policy and determine the corresponding ultimate reward score. This leads to high time complexity. we leverage the value function estimation method to overcome this limitation. This approach sets the iteration's search depth and employs the value function to compute the final reward. From a subjective perspective, a state where the drone is approaching the destination without any collision is considered to be better. Based on this, we utilize the estimation function shown in Equation (\ref{equ16}) for non-termination states.
    \begin{equation}
        \tilde{V}(s)=1-\frac{d(o, g)}{\max d(o, g)}, \quad \text { if } s \text { is non-terminal state }
        \label{equ16}
        \tag{13}
    \end{equation}
    The distance between the ownship and its goal is constant and equal to the diagonal length of the map. If there are no collisions with other drones or boundaries, the ownship receives a reward whose magnitude is determined by the distance between itself and the goal. Specifically, the closer the ownship gets to the goal, the higher the reward it receives.
    \item The process of updating the final reward and visit count for all traversed edges is called backpropagation. After reaching the termination state through the value function estimation function described earlier, the final reward and visit count for each traversed edge update. As the ownship traverses each edge, the edge accumulates a reward increment while counting the number of visits. The reward value for each edge can be obtained by dividing its accumulated reward by its visit count.
\end{enumerate}

 A single simulation consists of executing the four steps described earlier once. To improve decision accuracy, we perform a large number of simulations.
\subsection{XAI Framework}
In various UAV-assisted communication scenarios, in addition to meeting the key connection requirements for high-speed and stable data transmission, strict reliability requirements must also be imposed on UAV services. To address this, a scalable XAI framework is proposed in this section based on the characteristics of UAV trajectory optimization and collision avoidance, as shown in Fig.~\ref{System Model}. During flight, UAVs perceive their surroundings and take control measures. Real-time information about the UAV's environment is transmitted to both the XAI agent and the UAV flight controller. The XAI agent integrates scalable XAI methods to enhance confidence in decision-making for artificial intelligence systems. Both the environment information and XAI methods in the framework are scalable, with location and velocity information of UAVs and surrounding aircraft being part of the environment information and the MCTS method being used for XAI methods.

 Fig.~\ref{System Model} provides examples of questions that the UAV may raise, which can be answered by XAI. For service providers, XAI can improve wireless network service quality for heterogeneous users (such as mobile phones, PCs, vehicles, etc.) and enhance fault detection efficiency, with engineers being able to easily detect model decision-making errors. For individual users, XAI can provide details of flight decision-making to increase trust. For legal regulators, XAI can explain model decisions and establish trust in a quantifiable manner.

\section{Simulation result}
\label{section:Simulation result}
\subsection{UAV coordinate and power allocation}
To simulate UAV service coordinates and power allocation, we randomly distribute users within the service area and deploy the UAV near the initial height boundary of 100 meters. The UAV's flight range is 500 meters, with a width of 500 meters, and we employ a neural network with three layers and 40 hidden nodes. The activation function used is a rectified linear unit. The Adam optimizer is used to train the neural network.The greedy action strategy $\epsilon$ is set to linearly decrease from 0.9 to 0.1.\par
$\bullet$ \textbf{DQN}: The traditional Q-value coupled DQN inputs state information and outputs the action value of each action in this state.

$\bullet$ \textbf{Random}: The random method is a traditional approach for solving problems, which involves randomly choosing points within a specified area and computing the corresponding values at those points.

$\bullet$ \textbf{D3QN}: D3QN introduces double and dueling improvements, where the input of D3QN is state information and the output is the action value and advantage value of each action in this state.

\begin{figure}[h]
  \centering
  \vspace{-9pt}
  \includegraphics[width=1.0\linewidth]{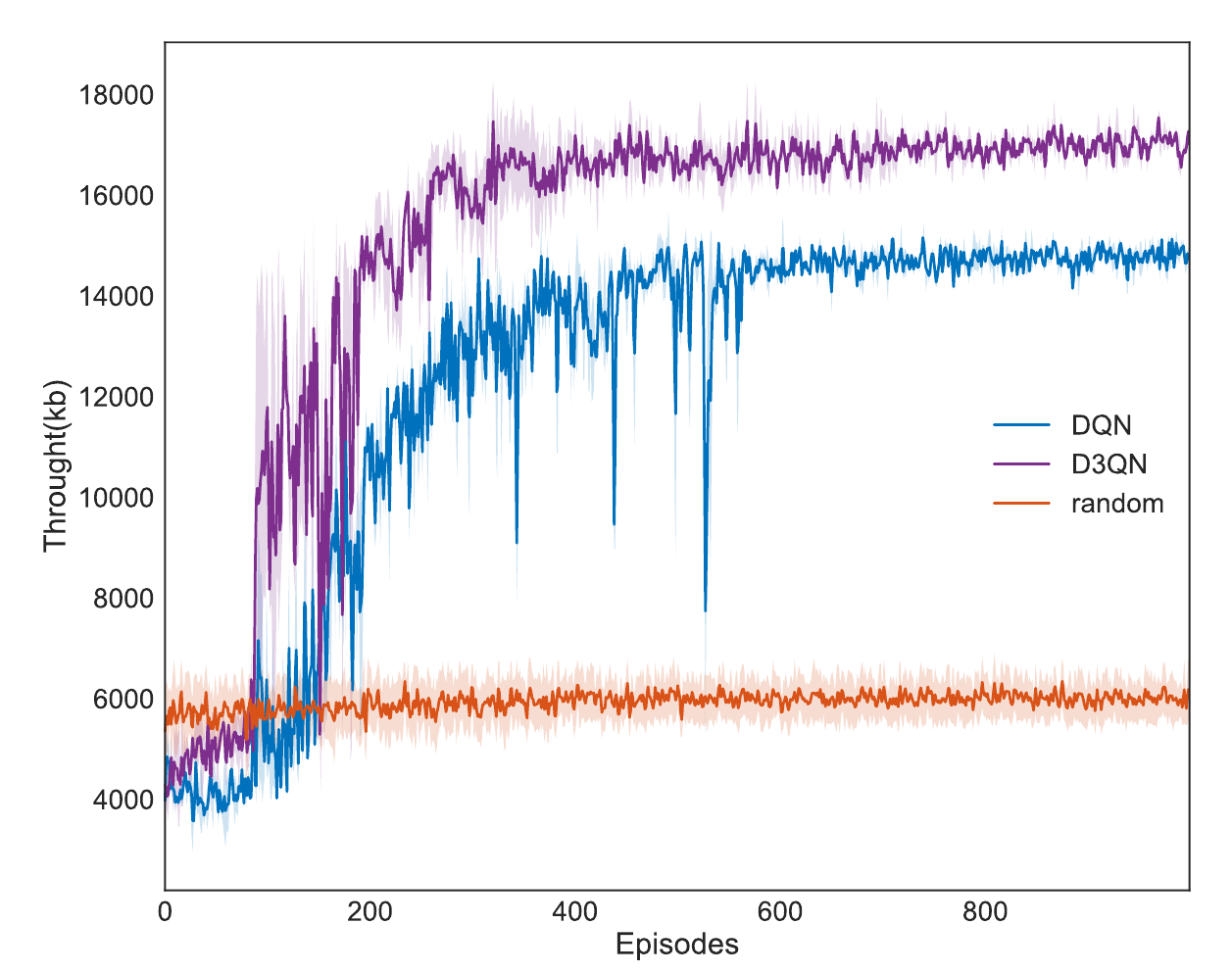}
   \vspace{-20pt}
  \caption {Convergence performance of different algorithm with $K=10$.}
  \label{covergence}
   \vspace{-5pt}
\end{figure}

Fig.~\ref{covergence} illustrates the convergence of the proposed D3QN algorithm. It can be observed that the D3QN algorithm requires approximately 300 episodes to converge, which is significantly less than the number of episodes required for the DQN algorithm to converge. Furthermore, Fig.~\ref{covergence} shows that the D3QN algorithm is able to converge to a performance of around 17000, which is significantly greater than the convergence value of approximately 14000 achieved by the DQN algorithm. Overall, the results presented in Fig.~\ref{covergence} demonstrate the superior convergence performance of the D3QN algorithm compared to the DQN algorithm and Random algorithm.
\subsection{trajectory optimization and collision avoidance}
In this section, we use the UAV service coordinates obtained in the previous section as the goal of the task. Intruders are randomly distributed within an area with a length and width of 2000 meters. reward is set according to Equation(\ref{equ17}), $d_{min}$ = 50.
 \begin{equation}
R(s)=\left\{
	\begin{aligned}
 1 ,& \text { if } s \text { is goal state }\\
  0.1 ,& \text { if } s \text { is time-out state }\\
0 ,& \text { if } s \text { is collision state }\\
	\end{aligned}
 \label{equ17}
 \tag{14}
	\right
	.
\end{equation}

$\bullet$ \textbf{DQN}: The traditional Q-value coupled DQN is trained in an environment with a fixed number of UAVs (using the states of all surrounding intruders as inputs)

$\bullet$ \textbf{Safe-DQN}: A safety-aware DQN model  consists of two DQNs: one ensures the UAV reaches its goal safely, while the other guarantees that the UAV does not collide with other intruders\cite{wang2022explainable}.

$\bullet$\textbf{Tree-fast}: A fast Monte Carlo Tree search method with low steps per iteration\cite{Yang2021}.

$\bullet$\textbf{Tree-depth}: Our proposed Monte Carlo Tree search method with a large number of steps per iteration and a search depth of 3 or 4.

\begin{figure}[h]
  \centering
  \includegraphics[width=1.0\linewidth,height=0.3\textwidth]{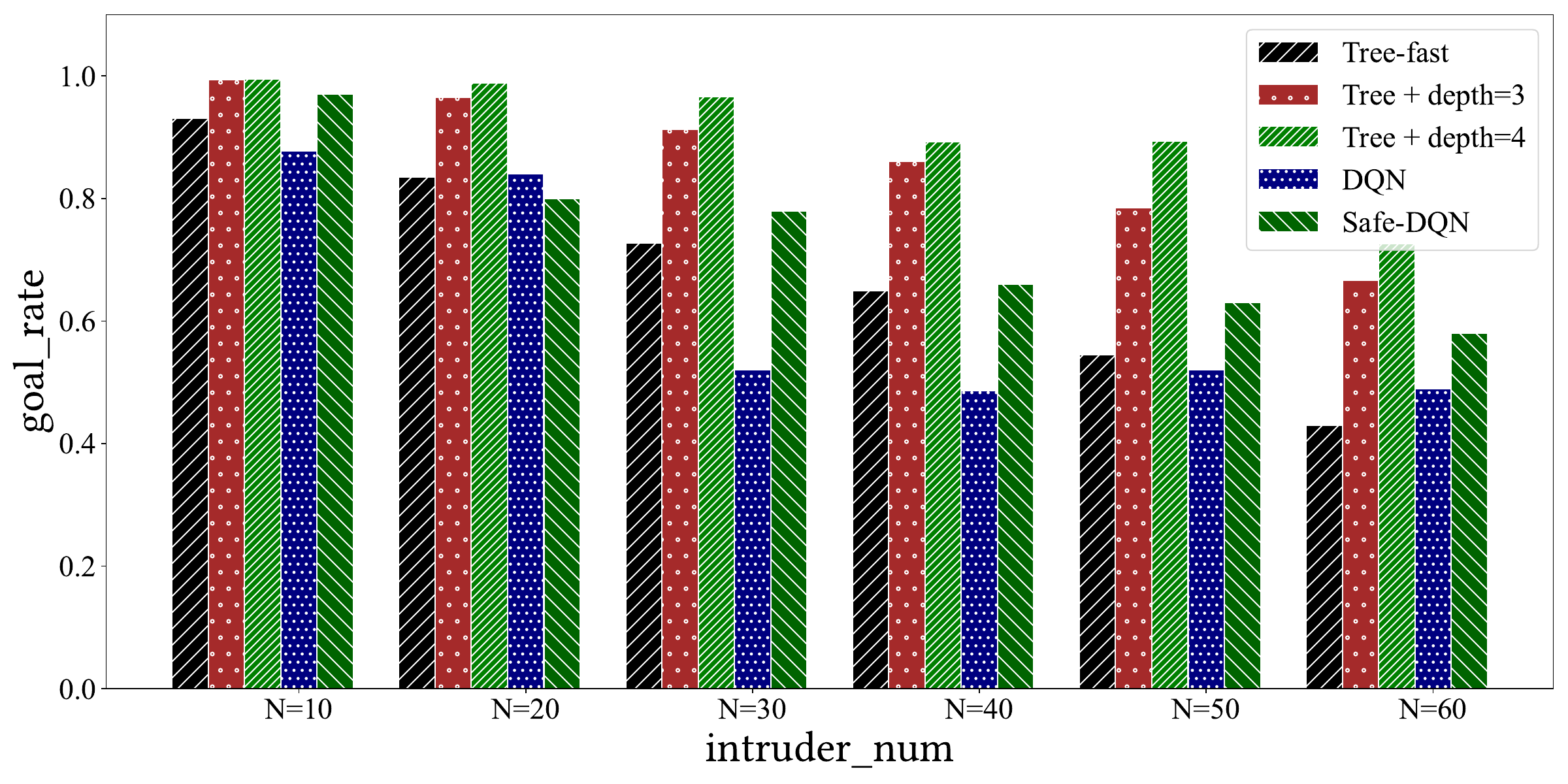}
   \vspace{-25pt}
  \caption {Performance goal rates of different algorithms with varying numbers of intruders.}
  \label{goalrate}
   \vspace{-5pt}
\end{figure}
\begin{figure}[h]
  \centering
  \includegraphics[width=1.0\linewidth,height=0.3\textwidth]{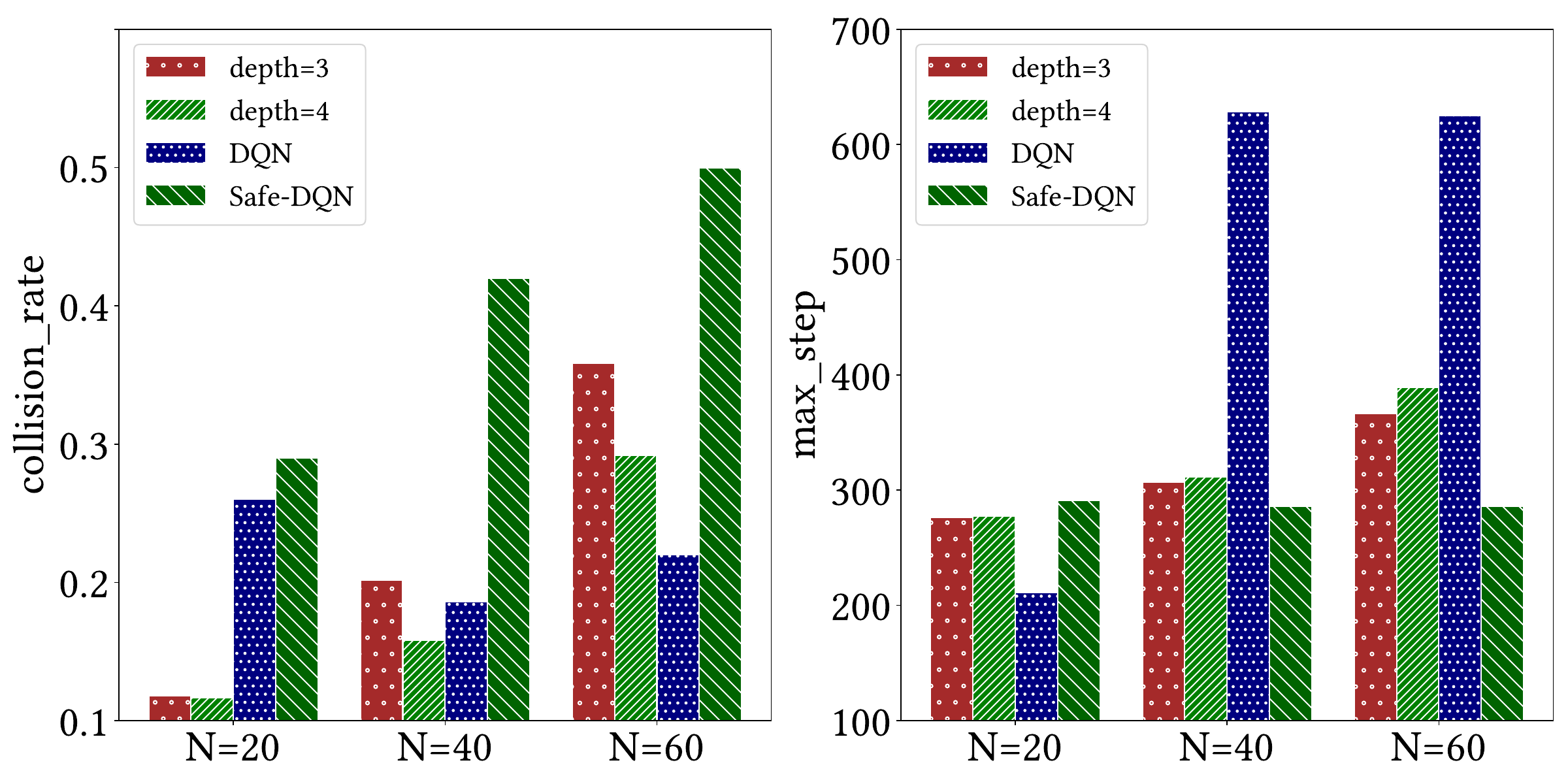}
   \vspace{-23pt}
  \caption {Collision rate and number of steps for different algorithms with varying numbers of intruders.}
  \label{two}
   \vspace{-5pt}
\end{figure}
We evaluated the performance of DQN, Safe-DQN, Tree-fast, and Tree-depth in trajectory optimization and collision avoidance under different intruder numbers, as depicted in Fig.~\ref{goalrate}. As the number of intruders increases, the MCTS algorithm can consistently sustain optimal performance compared to other methods. Additionally, as depicted in Fig.~\ref{two}, the MCTS algorithm maintains its overall performance while exhibiting a lower collision probability and shorter execution steps as compared to other algorithms. Notably, when the number of intruders increases, the algorithm's ability to generalize its performance is superior to other algorithms.

\section{Conclusion}
\label{section:Conclusion}
 This paper proposes an interpretable and secure trajectory optimization solution for UAV-assisted communication. It addresses the reliable UAV service problem by dividing it into two sub-problems. The first sub-problem is the constrained UAV coordinate and power allocation problem, which is solved using the D3QN method to determine appropriate UAV coordinates under spatial constraints, service quality constraints, and power constraints. The second sub-problem is the constrained UAV collision avoidance and trajectory optimization problem, which is addressed using the MCTS method to achieve reliable and secure UAV services. Additionally, we propose a scalable XAI framework, which achieves transparent and reliable decision-making during UAV collision avoidance and trajectory optimization processes. For future research, we aim to explore the application of interpretability in more complex scenarios such as the Internet of Vehicles and mobile communication.

\section*{Acknowledgement}
This work was supported by the National Key Research and Development Program of China (2020YFB1807700), the National Natural Science Foundation of China (NSFC) under Grant No. 62071356.
\ifCLASSOPTIONcaptionsoff
  \newpage
\fi

\bibliography{ref}

\begin{thebibliography}{10}
\providecommand{\url}[1]{#1}
\csname url@samestyle\endcsname
\providecommand{\newblock}{\relax}
\providecommand{\bibinfo}[2]{#2}
\providecommand{\BIBentrySTDinterwordspacing}{\spaceskip=0pt\relax}
\providecommand{\BIBentryALTinterwordstretchfactor}{4}
\providecommand{\BIBentryALTinterwordspacing}{\spaceskip=\fontdimen2\font plus
\BIBentryALTinterwordstretchfactor\fontdimen3\font minus
  \fontdimen4\font\relax}
\providecommand{\BIBforeignlanguage}[2]{{%
\expandafter\ifx\csname l@#1\endcsname\relax
\typeout{** WARNING: IEEEtran.bst: No hyphenation pattern has been}%
\typeout{** loaded for the language `#1'. Using the pattern for}%
\typeout{** the default language instead.}%
\else
\language=\csname l@#1\endcsname
\fi
#2}}
\providecommand{\BIBdecl}{\relax}
\BIBdecl

\bibitem{zhong2022}
R.~Zhong, X.~Liu, Y.~Liu, and Y.~Chen, ``Multi-agent reinforcement learning in
  noma-aided uav networks for cellular offloading,'' \emph{IEEE Transactions on
  Wireless Communications}, vol.~21, no.~3, pp. 1498--1512, 2022.

\bibitem{JSAC2018}
M.~T. Dabiri, S.~M.~S. Sadough, and M.~A. Khalighi, ``Channel modeling and
  parameter optimization for hovering uav-based free-space optical links,''
  \emph{IEEE Journal on Selected Areas in Communications}, vol.~36, no.~9, pp.
  2104--2113, 2018.

\bibitem{ZHAN}
C.~Zhan, Y.~Zeng, and R.~Zhang, ``Energy-efficient data collection in uav
  enabled wireless sensor network,'' \emph{IEEE Wireless Communications
  Letters}, vol.~7, no.~3, pp. 328--331, 2018.

\bibitem{IOTJ}
S.~Fu, Y.~Tang, Y.~Wu, N.~Zhang, H.~Gu, C.~Chen, and M.~Liu, ``Energy-efficient
  uav-enabled data collection via wireless charging: A reinforcement learning
  approach,'' \emph{IEEE Internet of Things Journal}, vol.~8, no.~12, pp.
  10\,209--10\,219, 2021.

\bibitem{TVT}
J.~Baek, S.~I. Han, and Y.~Han, ``Energy-efficient uav routing for wireless
  sensor networks,'' \emph{IEEE Transactions on Vehicular Technology}, vol.~69,
  no.~2, pp. 1741--1750, 2020.

\bibitem{TWC2019}
Y.~Zeng, J.~Xu, and R.~Zhang, ``Energy minimization for wireless communication
  with rotary-wing uav,'' \emph{IEEE Transactions on Wireless Communications},
  vol.~18, no.~4, pp. 2329--2345, 2019.

\bibitem{TWC2017}
Y.~Zeng and R.~Zhang, ``Energy-efficient uav communication with trajectory
  optimization,'' \emph{IEEE Transactions on Wireless Communications}, vol.~16,
  no.~6, pp. 3747--3760, 2017.

\bibitem{wang2019}
Z.~Wang, L.~Duan, and R.~Zhang, ``Adaptive deployment for uav-aided
  communication networks,'' \emph{IEEE Transactions on Wireless
  Communications}, vol.~18, no.~9, pp. 4531--4543, 2019.

\bibitem{Yang}
P.~Yang, X.~Cao, X.~Xi, W.~Du, Z.~Xiao, and D.~Wu, ``Three-dimensional
  continuous movement control of drone cells for energy-efficient communication
  coverage,'' \emph{IEEE Transactions on Vehicular Technology}, vol.~68, no.~7,
  pp. 6535--6546, 2019.

\bibitem{Zhang}
B.~Zhang, C.~H. Liu, J.~Tang, Z.~Xu, J.~Ma, and W.~Wang, ``Learning-based
  energy-efficient data collection by unmanned vehicles in smart cities,''
  \emph{IEEE Transactions on Industrial Informatics}, vol.~14, no.~4, pp.
  1666--1676, 2018.

\bibitem{Liu}
C.~H. Liu, Z.~Chen, and Y.~Zhan, ``Energy-efficient distributed mobile crowd
  sensing: A deep learning approach,'' \emph{IEEE Journal on Selected Areas in
  Communications}, vol.~37, no.~6, pp. 1262--1276, 2019.

\bibitem{Guo2020}
W.~Guo, ``Explainable artificial intelligence for 6g: Improving trust between
  human and machine,'' \emph{IEEE Communications Magazine}, vol.~58, no.~6, pp.
  39--45, 2020.

\bibitem{XAI1}
B.~R. Kiran, I.~Sobh, V.~Talpaert, P.~Mannion, A.~A.~A. Sallab, S.~Yogamani,
  and P.~Pérez, ``Deep reinforcement learning for autonomous driving: A
  survey,'' \emph{IEEE Transactions on Intelligent Transportation Systems},
  vol.~23, no.~6, pp. 4909--4926, 2022.

\bibitem{wang2022explainable}
L.~Wang, H.~Yang, Y.~Lin, S.~Yin, and Y.~Wu, ``Explainable and safe
  reinforcement learning for autonomous air mobility,'' 2022.

\bibitem{Yang2021}
X.~Yang and P.~Wei, ``Autonomous free flight operations in urban air mobility
  with computational guidance and collision avoidance,'' \emph{IEEE
  Transactions on Intelligent Transportation Systems}, vol.~22, no.~9, pp.
  5962--5975, 2021.

\end{thebibliography}
\bibliographystyle{IEEEtran}

\end{document}